\documentclass{article}
\usepackage{graphicx}
\usepackage{amsmath}
\begin{document}
\title{Organization and Complexity in a Nested Hierarchical Spin-Glass like Social Space}
\author{Fariel Shafee\footnote{current address fshafee@alum.mit.edu}\\ Department of Physics\\ Princeton University\\
Princeton, NJ 08540\\ USA.}
\section{abstract}
In this paper, we mathematically formulate the interaction and dynamics of a hierarchical complex social system where each agent is seen as an array of many variables.  The formation of identities and modifications can be studied by using interaction physics in an energy landscape.  The spin glass Hamiltonian is modified to suit our needs in light of complexity.  The expression of variables subject to weights and topology is considered.  Interactions within the same hierarchy level as well as the effect of one hierarchy level on another are studied. The effect of having many variables associated with a single agent brings about interesting dynamics.  The persistence of the identity units subject to stiffness, continuous changes and also sudden reorganizations is discussed.

\section{Introduction}
The problem of emergence and social networks has been recently become a topic of great interest. Various models have been proposed to predict the dynamics of many complex systems.  How the states of each member of a lattice may be updated by using simple rules depending on the states of neighboring members, eventually leading to complex patterns, has been studied (von Neumann 1966; Hopfield and Herz 1995).  Specific cases. such as the game of life (Berlekamp et. al. 2001), have been designed to somewhat mimic bio-systems. It has also been proposed that a new kind of scientific branch is required to to study the computational systems themselves that may give rise to complex patterns subject to simple programs (Wolfram 2002).

In another direction, the hierarchical nature of complex systems have been emphasized (Gell-Mann 1995, Ereshefsky 2000). Also, simple graphs of connecting agents have been successful to some extent in predicting the behavior of social networks (Newman et. al. 2006, Reichardt and White 2007).

In (Shafee 2009) we reviewed our work (Shafee 2004) where an agent as was defined as a multi-variable component of a connected hierarchical structure of identities. Previous models of interacting agents were mostly based on a single variable, eg. (Kawachia et. al. 2008).  Our work extended these notions to assimilate the more realistic complex nature of an individual. We modeled an agent as an array of variables in different states. These agents were then allowed to interact within a social and environmental context where other agents were placed as neighbors so bonds could be formed among agents.  Similar or complementary variable-states among agents caused the creation of affinity groups while dissimilar variable states created distance and animosity.  The result was an extended idea of {\it self} that allowed for the creation of groups and identity components at various levels with different degrees of stability.  The variables were also partitioned into genetic, skill, preference and belief types, and each variable-state was associated with a certain flipping energy, which was the cost of changing the state of the variable given its connection with others.   The interaction state-space for agents (each of them an array of connected variable states) placed in the social context, allowed for the formulation of an interaction energy landscape for the system.  The system itself opted to reach a minimum in the interaction space so that agents chose to align with neighbors, convert neighbors and reorganize the environment, or migrate to a coordinate more favorable for their existing state (Shafee and Steingo 2008).

In another paper (Shafee 2007), we tried to understand the nature of an agent in the context of a complex system.  This will be important when we formulate the variable space within the environment. An agent was modeled as a semi-closed entity so that he had a complex identity associated that he strived to preserve; however, the identity itself depended on being in contact with a larger environment.

In this paper, we lay a mathematical foundation for the interaction dynamics of such social systems, taking into account the properties of a complex social structure.  In our work, we make use of known fundamental properties complex hierarchical systems, and also characteristics basic to social organizations.  We add our own observations to create the space where interactions take place to predict the dynamics.

We first clarify the nature of the identity of an agent and formulate a variable space in conjunction with a physical space.  We then explain how interactions within the variable space are limited and augmented by the physical space.  We also describe the role of the interface of the agent and the environment given the agent's semi-closed identity.  Later, we formulate interactions within the space, taking into account restrictions posed by the the physical space on the variable-space in context with the agent's identity and vice versa.

Given the interaction-based nature of the system, we make use of tools used in condensed matter physics.
In the next section, we elaborate the idea of interaction among states in a complex system, and then modify the physical spin-glass model to accommodate features special to social and biological complexity.

\section{Specific Variables and Their Realization in a Complex Space}
\subsection{Review of a Complex System}
A complex system is special in the way that the configuration and the organization of the system enable it to recognize or process complicated patterns and/or behave in a non-trivial way, often reorganizing itself, in response to inputs, although the building blocks of the system may be simple.  An example is the brain, which has a neuron as a building block, and although the neurons individually performs a very simple task of receiving signals and then sending a signal based on whether the input reaches a threshold, when organized within the brain, a group of neurons can perform tasks as complex as cognition.  The human body itself is a complex structure made of cells that are organized into groups to perform special tasks and are linked together to coordinate many different functions.  The specialization of each body part and its link to other organs allow a human to coordinate different tasks and process information.

In a simple structure, interactions with the environment depend on specific parameters based on the statistical average of the properties of the constituent molecules.  For example, a rock can interact with another rock by gravitational interaction depending on the masses of the two.  When hit, the rock resists breaking down by applying a normal force.  However, the rock is unable to rearrange itself or modify any of its own constituents to preserve itself.  Also, if the rock is broken down into two parts, it maintains properties somewhat similar to those of the original one (scaled by the new masses) within each component.

\begin{figure}[ht!]
\begin{center}
\includegraphics[width=10cm]{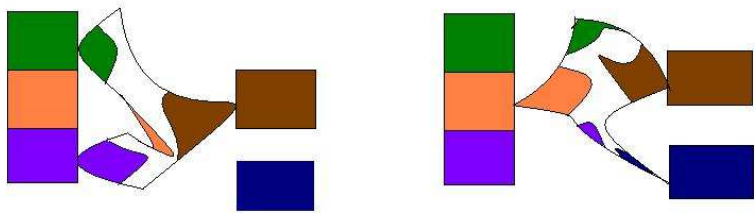}
\end{center}
\caption{\label{fig1} Two configurations of a complex agent in contact with different subsets of the environment states.  The rectangles represent dual environment states for agent states of the same color}
\end{figure}


\subsection{Functions of a Complex System}

In a complex system, the pattern for organization is important, and the function of the entire system depends on how the components are connected. The interconnection of many specialized components enables the system to acquire a multitude of properties and specializations for interacting with the environment.  The performance of the system depends on maintaining a coherent scheme for organization, which we can say is a pattern. Parts that can complement each other's survival can have an optimally connected coexistence in a complex system.  For example, in a human being, the property of visual recognition allows specialized cells to process light signals from another object to classify entities in the environment into different colors, sizes etc.  The process involves many rods and cones in the eye being stimulated chemically by different frequencies of light, and the signals from different classes (color recognizing) of cells being combined within a central processing unit to form an image.  Again, cells in the ear are specialized to receive vibrational signals within a certain frequency range.

The existence of many different types of blocks within a small physical space allows the system as a whole to respond to a large number of variables, categorize them, and also perform actions based on those perceived identities of interacting variables.

The occurrence of the diverse group of cells within a connected system that must act in unison creates an environment of cooperation and competition within that physical space. This may be somewhat similar to the environment within a spin glass system where frustration results from competing optimal alignments.  For example, a certain subgroup of cells may have a slightly different optimal pH than another subgroup.  However, since the cells are all connected within the same body, it may not be possible to satisfy the optimal environment for both, although both groups are needed to complement actions within the body for survival.  However, in order to interact with the environment, the connected system, ie. the whole body must act together, limiting the number of simultaneous actions.  The hierarchical order and the scale of components of the environmental states also necessitate a time scale and an energy scale for a task, restricting actions.

We let an array of infinitely many variables describe an agent in an abstract manner, which is then placed within the environment.  Of the possible agent variables, only a finite number are expressed within the environment because of the physical constraints imposed by the integrity of the system and the shared nature of the environment and the space within the connected system (Shafee 2007).  The variables again may be found in one of its possible states.  These states may indicate specific preferences, tastes, skill levels or markers (Shafee 2004).

For example, an agent may like to be in a yellow environment (which may be different from the liking of another agent).  Hence, the first agent has a preference for the state ``yellow'' of the variable ``color''.

Again, within a social cluster, two agents may have competing preferences for a variable-states but may share the same environment.  For example, one agent may want to paint the environment ``yellow''while another may like ``red'' and the two agents placed in a shared environment cannot be simultaneously completely satisfied giving rise to frustration-type mechanism.

In our paper, we define a variable space where agents interact based on the states of their own expressed variables.  However, the expression of the variables is constrained by the necessity for each agent to maintain his own interconnected and complementary variables' integrity within a physical confinement, and also the neighboring environment with which he is physically connected.  Hence, the array of an infinite number of variables associated with an agent is confined by a limited physical space within which the variables must be contained and interconnected, and the satisfaction of each variable is dependent on its ability to be connected to the other variables within the constraint.  The cost of a variable staying connected with other collaborative variables within the confined physical space may be its existence at a non-optimal, and hence a dissatisfied state.  Each variable component within the system tries to maximize its satisfaction by interacting with others while competing for the optimal environment in a shared space.  The system itself, consisting of the components, tries to maximize its satisfaction as a unit of identity at another hierarchy level.

We will be formulating these satisfaction terms as interaction terms between two states, where a low interaction energy indicates favorable alignment of neighbors and hence satisfaction or utility.

\section{Spin Glass and Our Model}

Our model for a frustrated social system is to some extent similar to a spin glass system, although the existence of complex entities, each with many variables, invariably makes the math more complicated, in the end requiring truncations and approximations to yield effective interactions.

We first review the basics of spin glass physics which will be relevant later in the paper.

\subsection{The Spin Glass Hamiltonian}

In the spin glass model, interaction energy is derived from the alignments of the electron spins and the coupling constant or exchange constant, J associated with each pair. When J is positive, similarly aligned spins are favored and when J is negative, opposite spins are favored.

The origin of J lies in quantum mechanics and the potential energy the electrons are subject to. The inhomogeneity of impurities in a spin glass system causes random distribution of J's, giving rise to frustration and making it impossible for the system to reach a global minimum steady state because of the coexistence of competing shared bonds.

The interactions among pairs may be short-ranged, restricted only between neighbors, (Edward and Anderson 1975) or long-ranged (Sherrington and Kirkpatrick 1975).

A typical spin glass Hamiltonian is given by

\begin{equation}
H=\sum-{J_{ij}\textbf{S}_{i}\textbf{S}_{j}}-\sum_i \textbf{S}_{i}.h
\end{equation}

Here, $S_i$ represents the spin of the i'th electron in a lattice.  h is the external magnetic field.

Although the spin glass model was derived to explain glassy systems in condensed matter physics, the properties are very similar to those in numerous complex systems, and the concepts of interactions among components allow one to associate similar Hamiltonians with many interactive systems (Bryngelson and Wolyenes 1987; Black and Scholes 1973).  The dynamics of such systems can be observed by noticing that a closed system tends to minimize the interaction potential.  Hence, energy landscapes based on designed spin glass like interaction potentials have been formulated and used to predict the dynamics of many complex phenomena (eg. Hopefield and Herz 1995).

\subsection{Spin Glass and Aging}
The mechanism of frustration gives rise to many local minima in the history of a glassy system.
If such a system is cooled below a certain temperature, given by $T_g$, it starts to undergo a process called aging, which involves fluctuations about meta-stable local minima followed by ``quake'' effects that irreversibly transform the system from one meta-stable minimum to another, reorganizing the entire system into a more stable configuration and resulting in a large release of energy (Sibani and Jensen 2004).  The quake effect is related to a record barrier.  An energy barrier is defined as the difference between the energy of the current state and that of the most stable (lowest energy) state observed in the history of the system. The record barrier is the largest such value observed so far. When such a record barrier is followed by an energy minimum record, the system enters a valley.  Hence, a spin glass system moves towards a new metastable configuration only when the new organization is more stable, which would mean that the new state is a lower valley in the energy landscape.

This property leads to the frequency of quakes getting lower as the system ages.  However, the system does not settle to an energetically optimal state in laboratory time, but progresses slowly. Within a small time interval, the system may appear to be in thermal equilibrium, displaying only fluctuations about the meta-stable state. Each quake reorganizes the entire system.  However, reaching a meta-stable state may take place following the system shifting from one sub-valley to another prior to finding the valley that is favorable for the entire system.  The decrease of quake frequencies in a time interval indicates the age of the system, which is the time elapsed from the onset of quenching.

\section{Dimensionality and stiffness of Spins in Social Model}
One of the main differences between the conventional spin glass model and the modified one appropriate for social evolution is the dimensionality of the variable space for each point.  Although models of social systems have taken into account individual decisions for the prediction of markets or trends (Black and Scholes 1973), human behavior and clustering are more complicated, and must include many affinity and choice factors, only a subset of which can be expressed at a time. As we mentioned in the beginning, the complexity of the organization of an agent allows for an almost infinite array of variables to reside connected within a physical confinement.  However, as was assumed in (Shafee 2009) some of these variable states may be fixed by genetic programming, some may have a high degree of stiffness (strong beliefs or social customs) and some variable-states may be acquired, and thus can be more easily changed. However, due to constrains explained later, only some of these variables can be expressed in interactions.

The notion of a flipping energy associated with variables in a complex context can be derived from the hierarchical connections of such variables, with some of these connections remaining in quenched or semi-quenched forms.

One way to model stiffness, and hence the required large amount of flipping energy to shift from one state to another for the case of fixed variables is to connect these variable-states with large environmental states that are held constant in a manner similar to a heat bath.

Again, the variable states defining an agent themselves are connected.  Hence, each agent himself may be seen as a spin-glass like semi-closed system, that is placed within a larger system consisting of more agents and the environment.  So though in a simple spin glass system, spins can be marked as $S_i$, indicating the spin state of the {\bf i'th} position of a lattice, in the social system, we denote the states by $S_i^a$, signifying the {\bf a'th} variable of the {\bf i'th} agents.  Within the agent himself, {\bf i} is constant and interactions take place among $S_i^a$ and $S_i^b$.  However, between agents, interactions take place among the states of the same variable (eg. whether the preference for color matches between two agents) and hence interactions are between terms of the forms $S_i^a$ and $S_j^{a'}$.

So the first major departure of our social model from the spin glass model is the existence of agents who themselves are meta-stable or semi-closed sub-valleys within a larger system.  Interactions take place among these agents who may not have constant positions with respect to other agents unlike in the case of a regular spin-glass model. Hence, interactions are among marked sub-valleys describing individual agents instead of between two indistinguishable electron spins based on their fixed spatial location.

\section{Mobility and Influence Space}
In a physical spin glass system, each electron is located at a fixed lattice point.  Its spin state is dictated by the states of the other electrons (for EA model, the nearest neighbors (Edward and Anderson 1975)) and the exchange constant for pairs of states for each specific location pair determined by the internal structure of the system.  The electron has the choice of preserving the state it has or flipping it, depending on the parameters, to move towards a more stable configuration.

However, for the case of a social system, each agent represents a complex configuration of states, which is somewhat different from that of another agent, although similar in many aspects.  The stiffness of the configuration depends on the internal couplings among states within the agent as well. Hence, because of internal couplings or genetic configurations, an agent may have stiff states different from those of other agents in the neighborhood.  In such a scenario, the subsystem, comprising of individual agents, have another option for moving to minimize unfavorable interactions in order to reach stability.  The clusters of individual agents, giving rise to social cluster identities can split into different clusters, or agents may opt to move to other clusters where interactions with neighboring agents are more favorable.

In (Shafee 2005), we have tried to formulate an influence space for interactions among agents. This is different from a fixed physical space for interaction.

The creation of clusters provide identities at a higher level of reorganization, forming meta-stable patterns with a group of agents so that interactions are highly concentrated among agents within the group.

In (Shafee 2009), such group identities were graphed by using an interaction energy landscape and by taking into account different types of variables that may lead to the formation of stable groups.
The splitting of one cluster, where interactions give rise to competing sub-domains into two isolated clusters may lead to the formation of stable patterns, which can be compared with events in history where tribes and countries split into two because of ideological differences (Cumings 1984).

Hence, the creation of clusters has an effect of partitioning the influence space in a hierarchical manner so that interactions take place between group identities that are allowed to form separately by permitting interactions among members from different groups to be minimized.

The agent himself has his variables connected within him in a small physical confinement.  He is connected with his nearest environment physically and also interacts with other agents that are his neighbors in an influence space.

The influence space is adjustable subject to constraints and stiffness, and defines identities of higher hierarchical levels such as social clusters and sub-clusters, which individually may exhibit characteristic statistical properties (Shafee and Steingo 2008).  However, the movement of an agent in an influence space, so that clusters can be reformed or split, depends on the stability of the cluster itself and the agent's cost for leaving the cluster.  A split of an energetically meta-stable cluster instigated by the departure of an agent may be resisted by the ``cluster identity'' although preferred by the individual agent. Again, the puzzling phenomenon of altruism, which prohibits an agent from leaving a social cluster even though this might come at the cost of the agent has been studied in light of the spin glass like model proposed by us (Shafee 2009b).
Hence it is the interconnection of different level of identities, which are meta-stable bundles of variables held together that come into play when controlling the dynamics of the system, allowing for components to get dissociated and move in certain cases.

\section{Overlapping and Nested Variables and Optimization in Each Variable}

In (Shafee 2007), we tried to understand the role of parameters in a hierarchical structure and the process of information leakage between hierarchy levels.  In another paper (Shafee 2007b), we have investigated the concept of states within the environment from the view-point of information.  We have also discussed the notion of states in a macroscopic world in (Shafee 2009d).

As was mentioned in (Shafee 2004), in our model, identities are fractal-like.  There is an agent, who is a a collection of variables that are held together within a semi-closed system.  Again, the agents form groups and clusters because of interactions among their variables, and with the environment (Shafee 2004; Shafee 2009b).  Hence, we can imagine levels of identities interacting and {\it nested} within different levels of a hierarchical system. In most cases, for a rational human being, the closely related variables within the agent, that depend on one another for survival, are expressed together as the ``semi-closed'' subsystem, giving rise to the typical economical prediction in the first order (Shafee 2004), making effects of affinities among variables less important.  However, the role of individual variable-identities may become important within a group from interaction among variable-identities.  An extreme case giving rise to altruistic behavior has been studied (Shafee 2009b).


\begin{figure}[ht!]
\begin{center}
\includegraphics[width=16cm]{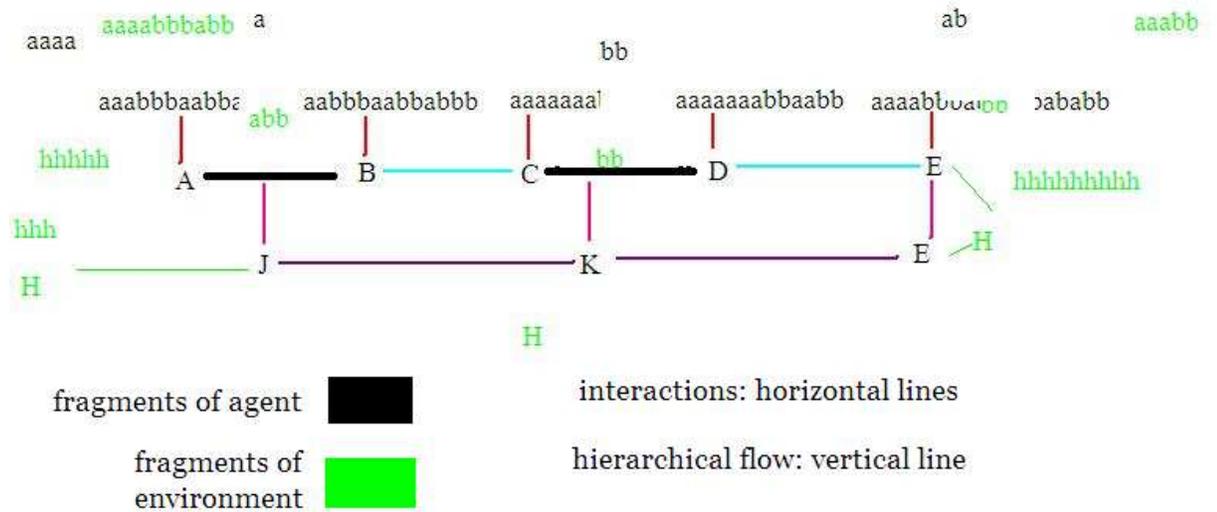}
\end{center}
\caption{\label{fig2} Hierarchical agent components interacting with various hierarchy levels of the environment}
\end{figure}

The concept of a certain variables at a hierarchical order emerging {\it from} interaction among levels within a nested hierarchical physical space may lead to many variables sharing the same physical space or be concentrated within a small physical space.

If we see a variable as a meta-stable sub-valley connected to other variable patterns within an agent in a dense manner, competition for shared segments within a finite space may arise from overlaps among portions of such entities.


Hence, a semi-closed complex agent may emerge from nested patterns that assume some degree of stability within an interconnected and overlapping framework.  Flipping the state of one variable, thus, may also imply partial change of an overlapped pattern, leading to its instability.  Similarly, two variable patterns may have a buffer region that may be optimal one variable-state pattern and not optimal for another.

\subsection{Interaction and Stiffness of Multiple Variables in a Nested Space}

The hierarchical nature of complex systems makes nested variables get expressed in different scales.  Some variables may be able to interact with several variables or influence a large number of variables (eg. environmental components effecting all agents within a region).

Again, if two variables are placed closely within each other's influence space, and one variable is in a larger and more stable position, the second variable coupled to it may have less influence in realigning the first larger variable into a more favorable state for itself.

However, if several similar identity units join in to form clusters of similar attributes as close neighbors with the first stiff variable, the combined interaction with the collaborating similar variables may change the state of the larger identity unit.

\subsection{Defining an Environmental Dual}
In our formulation, we model the variable space in such a manner that each variable-state has a possible environmental dual or specific range of environment that it can persist against.  While in a physical spin glass model, the spins can exist in one of two possible states, in the macroscopic complexity model, because of the hierarchical nature of states, and the possibility of states in many levels interacting, interactions may as well break down and disperse a state into lower hierarchical states irreversibly.  Again, stability may occur within a range of values. Although a proton must have exactly three quarks, a human being only has a possible range of height that is acceptable by the balancing constraints. Hence, a variable is assigned with a dual or a set of dual states with varying optimality within a range.

\begin{eqnarray}
S_x^a \sim h^a \\
S_x^a \sim h^{a \pm \Delta a}
\end{eqnarray}

with optimality

\begin{equation}
- J_{ih}^{a,a} S_x^a h^{a}
\end{equation}

when $J_{ih}^{a,a}$ have the lowest absolute values within the range of the dual, and approaches unfavorable outside the range.

\subsection{Optimization of a Variable}

Some of these nested variables have states fixed by genetics or environmental conditions.  However, in some cases, the optimality of a variable depends on the connection to other variables subject to local constraints.  Again, since sub-identities are nested, within a variable too, there might be islands of fixed attributes, while the sub-valley corresponding the variable is subject to interaction from the environmental and neighboring subcomponents.

The optimality of a variable depends on it being able to attain an energetic minimum subject to its internal constraints and its outside interactions.  In the case of an agent, that is semi-closed, we stated that the idea of individuality and a rational being derived from a large fraction of interactions being internal within the agent, such that a major portion of it is kept constant, while the environment is allowed to vary more within a tolerable range, but exhibiting a range of repeatable patterns in various combinations.

The complexity of an agent derives from a large number of variables interacting within the agent himself.  Hence, for the case of an individual variable within the agent, we restrict optimality by considering the fixed constraints within the subsystems of the variable itself, and its dependent couplings with other internal variables of the agent.  We keep the effect of the environment minimum to one dual environmental state in most of our studies for simplicity.

A variable that is not optimal may be in contact with a stiff environment state which is not its optimal dual, subject to its internal constraints prohibiting it from flipping to a state that is optimal with respect to the environment.  Couplings with other internal variables with different environmental optimal dual states and the restrictions posed by the agent's bundled identity might constraint actions from realigning the environment.  Hence, the variable itself might be subject to internal and external competing constraints, partially affecting the variable in conflicting states, making it non-optimal.

In the following expression, we show the different competing interactions leading to the optimality (interaction energy) of a certain variable.  Here, we are omitting all J's and are simply showing which states are interacting for readability.`

\begin{eqnarray}
H_{self}(^m S_i^a) = -^{m-1} S_i^{a_x} . ^{m-1}S_i^{a_y} -^m S_i^a . h^a -
^{m} S_i^a . ^{m} S_i^b - ^m S_i^a . ^{m} S_j^a -  [^{m-p} S_i^{a_u} . ^{m-p} h^{a_v}][h]
\end{eqnarray}

The superscripts to the left of a term denotes the hierarchy level.  The first term in RHS represents interaction among lower hierarchy variables within the variable, and the second term is the interaction with the dual environment.  The third term is the interaction of the variable with other variable states within the agent, the fourth is the inter-agent interaction along that variable.  The last term shows the effect of dissipation when the lower hierarchy constituents interact with components of the environment.

\subsection{Optimization in Each Variable and Fuzzy Limits}

The agent and the variable identities within the agent, each strives to reach a minimum but cannot, because of its interaction with competing neighbors, none of whom is an exact dual match for the entire agent comprising of many variables.  Each identity can optimize its own configuration energy by attempting to choose neighbors that provide stability within the pattern.  Similarly, each fixed variable, although genetically connected to some variables within the agent internally, and perhaps also to some fixed environmental variables, can also find topological configurations within the alignment space to avoid interaction with some other variables, or propagate/act to somewhat change its orientation/distance metric within the influence space.

We have previously discussed the notion of influence space and also ``turned off'' non-interacting states by choosing a set of expressed sets.


Now, an agent comes as a bundle of interconnected variables.  The interactions depend on the influence metric, which is also a function of physical distance.  Hence, if two agents share physical space in which they work, or which can be aligned along a state, then the array of variables of both agents are in close proximity in the physical space.  If the agents express their conflicting variables within the shared environment, the variables have terms of non-optimal interaction.  However, these terms can be avoided by turning states {\bf off}, and hence not acting based on these variable states.  However, the expression of a variable depends not only on a single agent or a group of agents placed in the environment, but also on the agent's connection with the environment and the other nested variables within himself.

A complex system by nature is made of many components and stability is subject to a range of conditions and scarcely discrete variables.  Although, optimal environments may have specific values for a state, usually a state is sustainable within a range, and optimality decreases gradually.  Hence, this arrangement makes it possible for several variable-states, each with slightly different optimal environments, to coexist in an interlocked manner if their mutual coupling is necessary for survival against other variables, such as dissipation and obtaining specific types of replenishment by interaction with certain other identity elements. However, such a tolerance towards marginally optimal conditions may also put a system towards breakdown subject to slight fluctuation.  The introduction of tolerable ranges built with thresholds in interactions is in a manner similar to using fuzzy logic.


\subsection{Preserving the Local Minimum}

The degree of complexity and the number of available variables in different states, together with the notion of an influence space capable of producing different degrees of freedom for each component's coordinates subject to its constraints, make an agent an individual, different from others.  Again, the agents may form higher order identities such as groups with varying degrees of stability.
In gist, an identity is associated with a fuzzy stable pattern found within the complex environment, where many of the building blocks of these patterns may be seen repeatedly in various combinations.  Each simple variable, that is part of a larger complex structure can flip states either discretely or continuously.  However, the interlocked nature of the agent with some fixed variables, lending to the idea of consistent histories that can be updated, associate a degree of stability or local minimum with any sustainable identity.  The longevity of such a complex identity-form depends on it being able to learn from the past to interact with a changing or unfavorable environment in manners (express the most favorable set of variables or adapt) so that a broad group of interlocked variable-patterns can propagate subject to finer changes needed for the preservation of the higher order structure.  The memories of past interactions teach the complex identity compound sequences of configurations that would yield the most stability.

The persistence of such interlocked sub-valleys, some in optimal alignments but some others in non-optimal alignments in a collaborative manner gives the overall structure a degree of stability and hence an overall identity that tries to preserve itself.

So with the concept of a complex identity, we assign the idea of a complex pattern consisting of many interlocked sub-patterns, that trying to preserve itself locally subject to competing interactions. As the complex system evolves, the interactions with neighboring structures also realign variables within the structure in the form of memory, making the interaction histories also part of the idea of self, and hence part of the evolved pattern, of such complex identities.

\begin{figure}[ht!]
\begin{center}
\includegraphics[width=10cm]{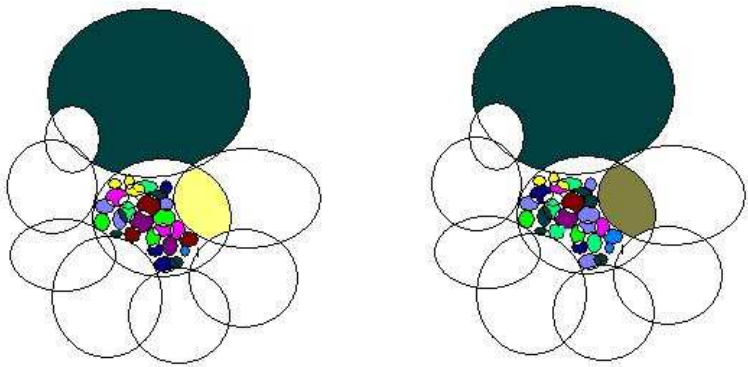}
\end{center}
\caption{\label{fig3} Changes in memory (small circles at the center) bringing about larger changes in connected preference states}
\end{figure}

\subsection{Environmental Dual Optimal Interactions}

A complex variables with interconnected variables also has a complex set of dual states attached to it, corresponding to its subsets of identities and interactions among them.

\begin{eqnarray}
^m S_i^a[=res \sum {^{m-1}S_i^{ax}} . ^{m-1}S_i^{a_y}] \\
\sim ^m h^a =[h^{a_1} . . . h^{a_n}]
\end{eqnarray}

The superscript to the left of a state implies the hierarchy level of the state given it is a nested expression derived from the residue of interactions among lower hierarchy terms (Shafee 2007).

Since the optimal environment has many components, the optimal environmental dual for two complex variable-states may or may not be null, meaning two complex variable states may share attributes in their dual. So it is possible to have
\begin{equation}
^m h^a \cap ^m h^b \neq \emptyset
\end{equation}

We now, from the point of practicality, view the agent as an entity that can move in a local environment and interact with states dispersed in an area. The complex agent and the environment are modeled so that the environment can display reachable states within constraints of agent's time and energy with a statistical property. Then the agent's optimality for being in that area depends on the statistical nature of available connections.  Hence, an environment with certain objects manifested within a geographical scale (which may be somewhat correlated, eg a desert is correlated with cactus), and surrounding properties created by specific arrangements of larger masses, eg air, weather and soil, may together be considered as an environment displaying certain properties that are within close influence distance of the agent.

Later, we will discuss the emergence of weights of variables because of the inter-connectivity of variables and the constraint in the number of total expressed variables at a certain time.  Here, we discuss another case of distribution among expressed/connected variables originating in the vast nature of the environment and the possible mobility in influence space.

The agent himself has many diversified variables interconnected within a limited physical confinement. This may require the need for a diverse array of dual environmental components to be present, which, in many cases may be impossible because of the scale of expression of environmental components.  However, ranges of permissible values make such complicated agent arrays of variables to coexist.

As we mentioned before, because of the scale of environment variables, and the definition of environment as a statistical ensemble that can be associated with a scale, a certain environmental state may bar the simultaneous existence of another, making it impossible for an agent to satisfy optimal connections for all of his variables, and optimizing one may come at the cost of maintaining another at a less optimal state.

Let us assume that an agent connects to the local environment statistically so that {\bf n} variable states are optimized.
The agent also incurs energy cost (remains in an non-optimized state) when shifting from one variable connection to another (traveling).
Now the average Hamiltonian in a time interval {\bf T} will be the sum of all optimized variables at different influence locations
and also the non-optimized total energy while shifting from one coordinate to another
normalized by {\bf T}.

Let the agent have the option of optimizing in {\bf n} or {\bf m} geographically distributed variables statistically, with time
{\bf $t_n$} or {\bf $t_m$} respectively the total time spent with these variables optimized, and $<E_{opt}>^n$ and $<E_{opt}>^m$
the respective average optimized energies. The statistical nature of the interaction also necessitates the agent to move from one point of interaction to another, accounting for non-optimized energy points within the total energy distribution  over time. The addition of extra variables in the statistical distribution of energy values in the time graph is preferred if the average interaction energy is decreased by adding the extra variable interactions.


\begin{eqnarray}
E_{ave}^n = {t_n <E_{opt}^n>+T_n <E_{nonopt}>^n}/(t_n+T) \\
\geq E_{ave}^m E_{ave}^m = {t_m <E_{opt}^m>+T_m <E_{nonopt}^m>}/(t_m+T)
\end{eqnarray}

However, the amplitudes of fluctuation of the data points may also come into play.  The roles of average energies and the amplitudes of fluctuations in the behavior of complex systems need to be studied further phenomenologically.


\begin{figure}[ht!]
\begin{center}
\includegraphics[width=10cm]{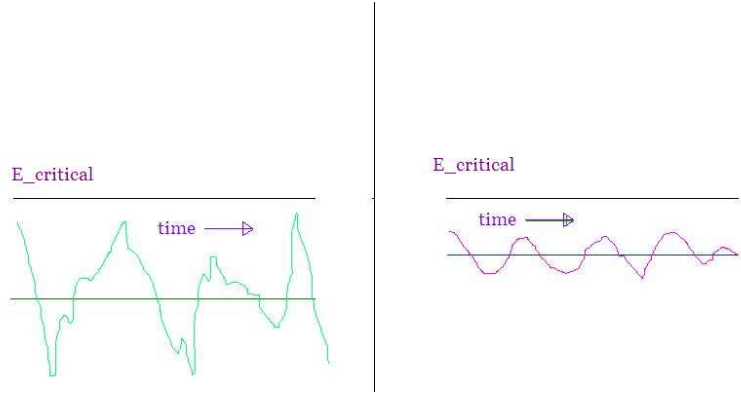}
\end{center}
\caption{\label{fig 4} 1. high fluctuation but low mean of interaction energy (high risk) 2. low fluctuation but high mean (risk averse)}
\end{figure}

\subsection{Environmental Dual States and Inter-agent Interactions}

 A variable that is part of a complex agent can have roughly three types of interactions.  The first one is that among other interconnected variables within the agent. In a complex agent showing a high degree of inter-connectivity, this gives rise to an identity making the entire connected structure able to interact with various aspects and properties of the environment. For example, the interconnection of the brain with the eyes and the ears enables the entire agent to respond to visual signals (how various objects within the environment interact with light) and auditory signals (vibration propagated through a medium within a certain range). The heart pumps blood to all the cells to make good for lost energy and to keep cells alive by repairing them.

We denote this type of internal connection with $H_{self}$.

However, as we said before, we also attach all variable-states with a dual environmental state, or a range of dual states that are tolerable. Interactions with environmental states are denoted with $H_{env}$.

The last type of interactions is among variable-states among agents.  As we just mentioned, these are passively fostered by existence and propagation of dual states and give rise to optimal bonds (friendship) or conflicting bonds (animosity).  We denote this category as

$H_{agent-agent}$.

The total contribution from interactions for a variable is the sum of all three types and is complicated by the existence of other interconnected variables within the agent.

Previously, we have given an expression for interactions between an agent's variable states and their environmental dual states.  Now we formulate how the existence of dual states creates bonds of similarity among agents.

In light of our previous formulation of dual environment states within a range of tolerability centered about $h^a$, again, omitting J's for clarity,

\begin{eqnarray}
-[S_j^a . h^{a + \Delta a}] S_i^a \le - h^{a + \Delta a} S_i^a\\
\rightarrow -[S_i^a . S_j^a] . h  \le -S_i^a . h
\end{eqnarray}

Here, the general h implies a vast environment which may contain the state $h^{a + \Delta a}$ statistically.

\section{Propagating Preserved Patterns}

The agent's idea of self is in the continuation of the variables subject to stiffness and dynamics (Shafee 2004).  The genetic pattern programmed into each cell largely dictates many variable-states that are held constant in the agent's life time.  When an agent loses cells, deforming that pattern, the agent's self preservation method, including processes like eating, finding shelter, tries to repair the lost cells to reproduce the patterns. Again, the neurons within the agent's brain cannot be regenerated, but the agent's idea of self includes the agent-environment interface in the form of memories by changes weights (spatial configuration in a large part) of the neurons within the brain (Shafee 2009b).

The environment also causes some genes to be triggered while some to remain dormant (Pospisil et. al. 2001), so that only some of the possible variables in the agent's characteristics to be expressed at a certain time.  Hence, the agent's idea of self or identity is in the propagation of a pattern of configuration of building blocks, which can be summed as a set of variable states in an abstract space.  The patterns have sub-patterns that are interconnected in a complex way and the stability of the pattern within an environment is dependent on the sub-patterns complementing one another in a semi-closed manner (Shafee 2007).

Although two agents at the same complexity level share a degree of similarity, the level of complexity makes each individual unique, especially since even in the case of genetically identical twins, two individuals are separate because of their different spatial histories, and hence memories that are parts of the evolving idea of self. The memory accumulated within each sub-valley during its evolution and history within the system, continually changes the configuration of the complexity space in the form of memory configuration, making any two such individuals distinguishable unlike in the case of electrons in a typical spin glass system.

Each such continually reorganized sub-valley of identity sees itself as a local minimum and tries to maintain itself within the larger environment.  The main difference between this model and the spin glass model in this context is that, in the process of aging in spin glass systems, the meta-stable states for the whole system reorganize irreversibly in jumps.  In our model, although we allow the entire system to undergo quakes in the form of social reorganizations, each individual is a local sub-valley which tends to preserve a local minimum as a preserved pattern that also undergoes continual irreversible change in the form of memory reconfiguration.  So the irreversible transformations for individual agent organizations have a continuous time scale (or a time scale that appears continuous to the agent from the point of view of the agent) for certain subcomponents. These continual irreversible slight changes in organization record memory of the evolution.  However, these are connected within a larger system, which may undergo reorganizations in bursts from one meta-stable valley to another.

Hence, we have a nested spin-glass type system where the identities are different hierarchy levels are connected at different time scales and spatial/influence scales as well, and it is the connection between hierarchy levels within one system, in a fractal type manner, that produces many layers of identities and complex dynamics.

\subsection{Sub-valleys in a Hierarchical Pattern with a Degree of Tolerance}

The formation of life and complex systems is subject to very narrow tolerance environments. In the entire solar system, so far only earth has been seen to have complex life formation.  On earth too, the stability of life forms depends on narrow environmental factors.  Even for the sustenance of the simplest possible life forms, such as a bacteria, specific pH ranges are necessary. Again, the advancement of existing patterns of life forms has been seen to proceed in stages (Darwin 1871).  One such theory of punctured equilibrium, which is similar to quakes in spin glass model (Eldredge and Gould 1972) show the split of species subject to isolation and environmental selection.

The ability for a complex life form to sustain its pattern in an evolving history depends on its ability to replenish lost components (environmental wear and tear, otherwise deformation of pattern because of interaction with levels of intolerance).
The repair mechanism involves having an ecosystem where one life form can use components from another life form (for example, protein from another life form available instead of having to synthesize complex molecules using carbon, nitrogen and hydrogen molecules) in an energy efficient manner.

While in a regular physical spin glass, the J's are fixed at spatial locations causing identical electrons to undergo configurational change from one meta-stable system state to another only when energy supplied to the system is greater than the record barrier, in the social spin like model, the fixed J's are local to patterns that are allowed some degree of freedom in an influence space.  Hence, bringing one pattern as the neighboring environment of another agent causes the agent connection to create a combined system that may update the configuration of the agent.

This property is derived from the agent's identity depending on the internal couplings in the form of quenched J's and also the extended interactions with the neighboring environment, which is not constant, but may introduce patterns with different distributions of J's within itself that now also form bonds with the agent's states.

Hence, the model of social spin-like systems involves having an environment with repeating patterns in many degrees of complexity and the probability of a more complex pattern being created depends on less complex structures with similar building blocks existing.  The relative meta-stability of a complex structure derives from being able to preserve its pattern by interacting in a more or less homogenous greater environment subject to a tolerable degree of variation, which contain patterns in different levels of hierarchy.  If the loss of a complex part in an interaction of a relatively disordered environmental configuration can be re-established by interacting with structures with somewhat similar components, so that interaction with another pattern at the interface of the agent-environment, causes the agent's former pattern to be recreated within a certain degree of accuracy, the pattern can gain stability in a nested spin glass like system where similar local minima in the form of sub-valleys might recur and interact to give stability to more and more complex patterns in configuration. The stability of a complex pattern depends on a narrow range of hierarchical patterns existing, making a tolerable environment with possible histories subject to rules in the form of repeating histories of interactions at different hierarchical stages.

\section{Expressed and Suppressed Variables}

As was discussed before, the agent, due to its complex nature, can be approximated to have an infinite number of variables that can possibly rise from all types of interactions among various levels of identities formed among its building blocks.
The level of complexity assigned to an agent is with resect to its neighboring structures and forms of identities, and is not an absolute standard.

When an agent is placed in an environment, components of the environment may affect various attributes of the agent.  An agent can perform a task, thus interact with the environment, to express a certain variable-state within the agent in the form of molding the environment along that preference.  Hence, an agent's variable-state may get expressed within the environment in the form of the effect it has on the environment due to interaction.  For example, an agent who prefers the color red can paint his room red, expressing the variable.

However, as mentioned before, a complex agent may have infinitely many such preferences, all of which cannot be realized because he cannot carry out all types of actions simultaneously (shared physical realization) and also because an action takes time (Hence, we can introduce a coarse grained time scale with interaction for different types of variables).  Different variables thus come with different costs.
Some take more time and energy to change the environment, and hence get expressed.  This cost is included in the flipping energy (stiffness) of the corresponding environmental variable.

We can try to formulate this by imagining a buffer state $S_i^{{buff}_{a/b}}$ that can act as a buffer between the environment and either $S_i^a$ or $S_i^b$ by being oriented along $S_i^{{buff}_a}$ or $S_i^{{buff}_b}$.  The resultant agent/environment interactions can be formulated by either of the two expressions below:

\begin{eqnarray}
-.  S_i^a . ([S_i^{{buff}_b}]S_i^b) . h^b =E(b_{expressed})\\
-. S_i^b . ([S_i^{{buff}_a}]S_i^a) . h^a =E(a_{expressed})
\end{eqnarray}

Here, the E's represent the energies corresponding to the Hamiltonians when either a or b is expressed.

The expression of a variable onto another system derives from the variable being able to interact with and influence the system to bring about some change or reorganization within the system or by adding some constraint on the system.

A variable that has no influence on a system is unexpressed in that system.  When a semi-closed complex agent interacts with the environment, the interactions depend to some extent on the constraints imposed as mentioned before.  Hence, not all variables that can be used to define an agent are expressed.  An agent can choose to do certain things at the cost of $\it not$ doing certain other things. Again, if two optimal variables are interacting, the coupling imposes a constraint on the variables from shifting or switching to other states because of the local energy minimum created by that interaction.

Now, again, a change is brought about to a certain state by its interaction with another variable state only when the first state connected with it can exchange enough energy to bring about the change in configuration at the variable level.  This process may require collaborative effort from different coherently organized components from one or more agents, or prolonged leakage of energy/influence to slowly change portions of the interlocked patterns until a threshold is reached so that the entire variable state makes a transition.


\begin{figure}[ht!]
\begin{center}
\includegraphics[width=10cm]{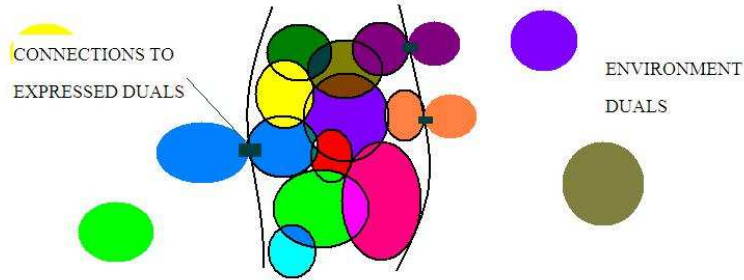}
\end{center}
\caption{\label{fig5} The effect of placing an agent with certain states in an environment}
\end{figure}

The degree of interaction and the level of change produced/constraint imposed by that interaction is indicative of the level of expression of a variable.

\subsection{Evolution of Interactions}

One system inspired by the spin glass model is that of associative memory in the context of neural networks. The modification of spin glasses in this context involve the introduction of weights instead of varying J's. Each neuron can be in +1 or a -1 state and similar states always produce a favorable energy in the energy landscape (Hopfield and Herz 1995).
Two patterns are matched when most of their neuron states are the same.  However,
in this model, each neuron pair is weighted, dictating what fraction of one neuron's signal influences another.


Our social model incorporates the idea of an influence space as well as the concept of compatibility between states giving rise to different degrees of exchange energy among interactions between different states that lead to the production of some meta-stable patterns.

Hence in our model, we incorporate the idea of J in the variable space as the affinity of one state for another state, while also keeping an id  ea of weight that indicates the expression or degree of connection between two states because of configuration.

The notion of weight is important because, as we mentioned in the beginning, although two interactions may be favorable, the restriction of the physical space shared by state and because of coarse graining in interactions at a hierarchy level, one set of interactions may hinder another set making one variable expressed while suppressing the other set of favorable interactions.

In any configuration of the physical space associated with the variable space, weights associated with different interactions are readjusted to accommodate constraints posed by sharing, physical limits, coarse graining and time.  The reorganization of weights can be mapped into different topological deformations of the physical space with which the variables are connected in the presence of specific environments and conditions because of internal physical connections between variables.

\subsection{Weight and Stiffness}

In the beginning of the paper, we discussed the existence of variable states with different degrees of stiffness.  We mentioned the role of genetically fixed variables, beliefs, acquired tastes, preferences and skills (Shafee 2004).  In the previous sections we also discussed the narrow range of environmental variation and the role of hierarchical nested local sub-valleys exhibiting repeatable pattern components.  In the contexts mentioned, we now derive the existence of different degrees of stiffness of variables.

In the abstract variable space, we can introduce some environmental fields, denoted by $h^a$, that interact with and influence specific agent variables $S^a$.  In the context of the narrow range of environmental tolerance for complex systems, we can keep these specific fields constant or may separate them into a few broad categories throughout large chunks of the environment.  These fields are coupled with specific agent variables with a high weight in the agent-environment Hamiltonian.  These are variables that are programmed to specific states by the environment, and are hence genetically fixed as common qualities.
Some variables may be varying from agent to agent, but fixed within an agent, giving the agent his individual attributes.  These varying but fixed attributes may be all within the range of tolerance of the environment, but the agent's gene has one of the many possible degenerate or nearly optimal values programmed permanently. In that case, each agent has a specific set of $[J_{ih}^a]$ programmed for a given set of {\bf a}'s that represent possible sets of the environment. These variables may come in different weights, dictating the importance of such variables in a social interaction landscape.

On the other hand, social norms and traditions may have a large degree of stiffness because of their connections with a large number of similar variable states in other agents who are also somewhat similar with respect to other variable states within a cluster.  Repeated positive interactions leading to optimal interactions and satisfied states prompt an agent to maintain such connections within a cluster and identify strongly with a group that may share similar norms.  An agent's identity derived from both past history (Shafee and Steingo 2008) and present affiliation groups may cause a certain norm or tradition state to have a large number of connections yielding optimal interactions.  Hence the stiffness of a certain variable state may also derive from an agent's security from interacting with a group of agents sharing similar attributes in other variables.  The notion of such security of shared information within ethnic groups have been experimentally studied (Bowles and Gintis 2004).

\subsection{Stiffness and Hierarchy}
In the physical spin glass model, two electrons have similar influences on one another. The exchange term derives from quantum mechanical properties of two electrons and the potential they experience.  The interactions between the magnetic moments of two electrons are negligible.  However, when an external magnetic field is applied, electrons tend to align along it (The $h.S_i$ term).  In a social model depending on hierarchical structures, intermediate interactions between levels exist where the influence of one level on the other is asymmetric, though not always as determined as that between an external field and a small electron.  An agent experiences some components of the environment as stiff aligned fields, while he can also interact with and change his local environment. Such asymmetric properties of interactions can give rise to asymmetry in stiffness, and hence the $\bf J$ factor among variables at different levels.  The effect of coherent action of agents against the environment has been discussed.

\subsection{Weight in the Nested Landscape}
If we take a statistical ensemble of expressed variables over time, the amount of expression may be indicative of the weights of each variables. While an agent strives to align the environment along his highly weighted priorities in a long term basis, low weight terms may not have sufficient interaction with the environment to be expressed at a macro level to be discernible.

If we simply consider the effects of variables or total interaction of variables in a duration, we can imagine a statistical ensemble

\begin{equation}
w_1<S_1>+w_2<S_2>+.........
\end{equation}

The main difference between the weight and {\bf J} is that while the {\bf J}'s are specific to interaction for a pair of states, the weights are subject to the total configuration of the system and the environment, and hence the priority of a specific state being expressed within a system. This priority may depend on the {\bf J} between a pair among many other factors.

Now, let us model the effect of weight in a simple manner.  As we mentioned previously, the variables exist in an interlocked manner within the complex agent, and not all variables can get expressed.  For simplicity, let us assign an interaction surface for the whole agent, $A_i$ and an interaction surface needed for the expression of a variable, $A_a$.  Hence, the cost in common resources shared with other variables can be put in a fraction, $A_a/A_i$.  The other important cost comes from the competition of the variable-state's identity and the agent's identity, ie. the cost incurred by the interlocked network in the energy landscape when a certain variable state is expressed against the optimality gained by the variable itself from being expressed (ie. by aligning the environment in its own direction, which is dependent on the coupling constant, $J_{ih}^a$).  If the agent in its higher order identity experiences a high cost itself by allowing a certain variable state to be expressed, the other variable-states within the agent will suppress the first variable state.  Also, some variable states may be critical for the existence of the agent, and hence must be expressed to keep that agent intact.  Hence we can say that in our simple model, the weight is proportional to

\begin{equation}
(\Delta H_{variable} -\Delta H_{agent})/(A_a/A_i)
\end{equation}

Here, the $\Delta$'s express the change in energy when the certain variable-state is allowed to expressed.

\subsection{Flow of Energy and changes in the Social Model}

In the physical spin glass model, a record barrier must be reached prior to an energy minimum for the system to reorganize irreversibly.  Otherwise, the system fluctuates about the values of the meta-stable valley.

Our social model is inspired by such spin glass systems and their evolutions, but has major differences because of the in-built nature of social systems.  The first difference was already pointed out when nested hierarchical patterns with repeatable components and a narrow degree of tolerance within the environment were mentioned.  Hence, we are talking about a very specific type of system that exhibits interactions among patterns that can be formed, sustained as semi-stable components within the system, and can interact in a very complex nested manner, with fractal-type identities being expressed.

The idea of local stability given semi-openness with neighbors asks for modifications, constraints and new types of dynamics.

We have already discussed the notion of a flipping energy required to flip a state in a complex system.  The semi-closed nature of complex hierarchical systems makes these flips somewhat different from the flip of an individual electron spin or the quake associated to the whole system.  These flips are somewhere in between, since the subsystems exist as identities within a bigger environment, and also have components within themselves.  Hence a flip of a variable necessitates reorganization within the variable, but does not necessarily have to be irreversible like a quake since the variables are not closed, but in touch with other variables in an environment.
However, a degree of stiffness is associated to changing the state back because the reverted state must be energetically more stable within the larger system than the new state because of the new environmental configurations.  In spin glass systems, rejuvenation can happen because of temperature fluctuations (Sibani and Jensen 2004).

Reorganizing an identity requires breaking older bonds and frozen configurations, and hence elevating the system to an energetically unfavorable position prior to settling in a record minimum.

In the spin glass system, energy can be supplied in many ways.  One of the methods is applying an external magnetic field.  In the social model, as was mentioned previously, a state can be associated with a flipping energy required to make a transition from one state to another.  In a physical macroscopic world, transformation of a system is by the means of work, which can be provided by the free energy of a system.  When an agent works to change the state or organization of any object, he transfers free energy.  The amount of work dictates the change made. Usable energy in the form of work can change an entire system together, reorganize it or move it.  Energy is also lost in the form of dissipation, as parts of systems disperse into the environment or provide degrees of freedom to the environment in the form of heat.

Hence, usable work can be seen as the effect of a variable at one hierarchy level influencing another
variable at a similar level.  Lost energy in the form of dissipation takes place because of interactions among subcomponents at lower hierarchy levels that are lost from the larger identity.

The effect of the states in our model themselves being complex entities consisting of lower hierarchy states comes into play when smaller variable states interact with larger ones. For example, if three repeating patterns are locked in an optimal configuration, and the fourth pattern from a neighboring agent or environment is not-aligned, realigning the fourth along the first three may be optimal for the sub-system made of four patterns since the resulting field from the three patterns has a larger effect.  However, fixed patterns or stiff patterns held by genetics or environmental stiffness
create competition by elevating the flipping energy.

Hence, in gist, to express the series of events taking place in these variables trying to align one another along their own preferred direction subject to stiffness, we succinctly assign a flipping energy (required for a variable to flip rom one state to another) and also flipping energy density (in time and space) that two connected variables send to one another to align the other along its own preferred direction to optimize its own stability.  A similar model with axioms was proposed (Shafee 2002).

Mathematically, we can simplify a complicated process of slowly changing a state by repeated interaction at a steady rate of coherent actions by multiple interacting states by saying that the coupling energy, J, is increasing with time and number of interacting agents, so that when the cumulative J reaches a threshold, the overall macro state is seen to change to a resultant optimal condition.

\begin{equation}
J_{jx}^a = \sum_x \int J_{jx} dt
\end{equation}


\begin{figure}[ht!]
\begin{center}
\includegraphics[width=10cm]{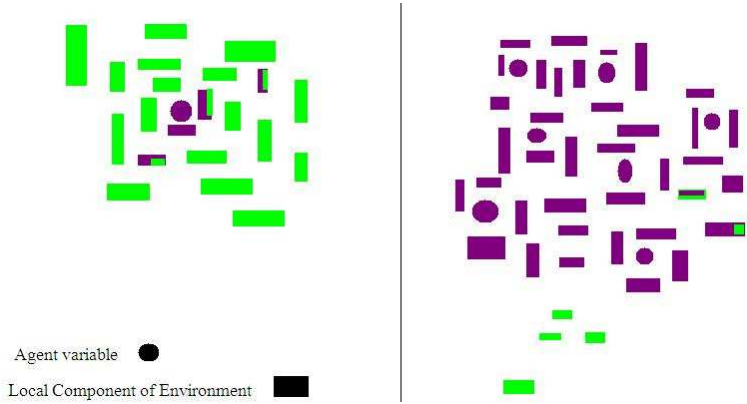}
\end{center}
\caption{\label{fig6} The effect of placing few and many coherently aligned agents in contact with a larger environment state}
\end{figure}

In a physical spin glass model, the flips of a spin are instantaneous.  However, in our social model, where variables are macroscopic, the states can change more slowly, and a mismatch in the rate of change in different states can come into play.

In (Shafee 2009d), changes in a detector state based on interactions with quantum objects were described.  The final indication of a state in a detector included reorganization of many components leading to a shift in the state of the final object.  In the macroscopic world, within states, properties are manifested based on their ensemble averages and in some cases sums (eg. for total mass or charge), and interaction among objects are based on these properties displayed at the macroscopic object level.  However, in a hierarchical environment, components also interact with other smaller components of the environment (eg air molecules) that are made of the similar building blocks such as neutrons and protons, that can also interact with objects at a certain level, leading to dissipation.  Free energy involves a macroscopic object being able to retain its macro properties by propagating that pattern or maintaining it when placed with other variables from different hierarchy levels.  The macroscopic reorganization of states includes many smaller objects, some working in unison, with various time-scales, making the the concept of energy transfer rates among states a relevant issue.

Hence, although in the variable picture, interacting variables can be formed in the spin0glass like manner, the macroscopic nature of these variable states and the hierarchical structure of identities make it necessary to include the notion of rates and time-scales.

\subsection{The Notion of Temperature}

The notion of temperature is related to the average energy per degree of freedom for a molecule.  If a molecule has many degrees of freedom, more energy is required to raise its temperature by a certain amount.  Similarly, since temperature measures average energy per degree of freedom, if molecules of different masses have the same temperature, the more massive ones display a lower speed.

Again, temperature is related with entropy by the following equation:

\begin{equation}
T = {\delta E}\ {\delta S}
\end{equation}

Entropy is a measure of disorder in a system.

In a viable social complex model, we have repeating building blocks and semi-stable patterns.  The same patterns are expressed in different permutations and combinations subject to possible stability conditions within a tolerable narrow range of environmental conditions.

In our model, we allow some of the states to be flipped.  The flipping is subject to a flipping energy, and the number of connections may make some states stiff or massive.  It is possible to assign the notion of entropy in the social context as an indicator of diversity of states in an ensemble.  The notion of temperature may similarly be associated with energy per degree of freedom.  A high temperature would thus indicate the possibility of more excited states. However, these require further modeling, and are not essential for the purpose of understanding our basic model that produces dynamics based on interaction and frustration originating from the various {\bf J}'s.

So in this paper, we simply discuss the process of interaction within components of a system and dynamics arising from competing behavior among semi-stable interacting patterns.


\section{Interactions and Survival}

\subsection{Critical energy and Local Minima}

The interconnected nested nature of variables in our hierarchical complexity model, the competition among states within the complex identities given the necessity to adhere together because of interaction optimality, and also the possible interactions with random elements of the environment causing dissipation (Shafee 2007) make it necessary to include a threshold in our model.
To simplify the complicated process of dispersion and the breakdown of a configuration, we assign a critical interaction energy $H_a^{crit}$ that is the cutoff for the stability of the configuration-pattern.  If the components are not able to interact within the threshold minimum, competing interactions with random components of the environment disperse the variables into smaller hierarchy levels and destroy the pattern irreversibly.  This is the death of an agent.  As was proposed in (Shafee 2002), the inclination of an agent to preserve his overall pattern is part of the social system in a tautological manner since if the agent did not have the quality of preserving its identity, the identity would simply not be preserved.
Hence, within this specific system, the agent must have an in-built mechanism or structure that tries to preserve its overall identity by maintaining the threshold energy.
Making the Hamiltonian exceed the threshold introduces an irreversible process of breakdown and losing the components.

\begin{equation}
E(interactions)< E(cutoff)
\end{equation}

Here, the interactions are at the level of the variables of the hierarchy level of the identity structure, so that dispersion leading to the loss of the expression of macro-states at the highest hierarchy levels, leading to interactions at sub-levels of building blocks that make the agent components indistinguishable from other environment components, causes to the breakdown of the complex structure capable of expressing variable states emerging from various levels of hierarchy and interactions among those specific to that entity.  Again, the destruction of an identity structure may or may not cause substantial damage to another identity structure at a higher hierarchy level of which it is a component, depending on the weight of the lost identity.

The role of critical connections within an agent can be understood by means of variable-states that are connected optimally with many other states so that removing the first brings about a large change, often leading to loss of the complex identity, or vital connections that can be assigned a large $J_i^{ab}$.




However, as was mentioned in (Shafee 2004) all of the components within the agent also have their individual identities, and although in out interaction based model the agent's individuality comes in as the constraint for the behavior of a rational being, the formation of higher order identities such as groups resulting from the interaction among different variables within agents also impose such survival constraints of these identity structures subject to the optimal couplings, and the degree of match, making the agent a component of a larger meta-stable structure. The couplings between agents based on similar variables may reach a critical level such that when the identity of a larger cluster of agents and the identity of the agent are competing, there may be instances in which interactions among similar variables between agents may cause an agent to sacrifice himself for the perpetuation of the cluster (Shafee 2009b).

\subsection{Environmental Changes: Adaptation and Mutation}
In the previous sections we mentioned the stiffness posed by genetic programming or social connections needed to maintain certain constant or slowly varying configurations within an agent that help an individual maintain his identity throughout lifetime.  The interactions among these fixed states within a social cluster may often be complementary, eg. in terms of talents, which may foster the creation of higher level organizations such as social clusters that depend on similar as well as complementary variable states among agents.  In (Shafee 2009) graphs were drawn regarding such organizations.

However, each agent is also connected with his local environment, which influences the optimal state of his variables.  The notion of punctured equilibrium is well-known in evolutionary theory (Eldredge and GOuld 1972).  Recently (Anderson et. al. 2004), a study has been made using the total number of people in a population as the record variable to observe the effect of mutation in moving a species into another in a manner similar to quakes.  A gene pool was created which increased in diversity with an increasing population,
and random agents were allowed to die probabilistically, while interactions among agents with different genetic materials led to asexual birth of agents in the model.  The study displayed the existence of spin-glass-like behavior in population dynamics.

In that paper, the reason for using population size as the record was not clear. the theory of punctured equilibrium in evolution maintains that two conditions needed for
the split of species are isolation and of population (Mayr 1954), which were not taken
into account in that paper.

The formation of social clusters and different degrees of identities because of interactions were not discussed.

Our model looks at the connection between the environment and interactions among variables more closely, and in the context where the creation of a new species is not necessary.  As the local environment changes, the agent variables may flip states to align it with the environment (adaptation), or the variable space may undergo reorganization to express a different set of variables (adjusting priorities and the existence of vestibule organs).  However, if the variable states are genetically or traditionally fixed and highly weighted, the agent may be left in a dissatisfied state. Hence, in our model, the effects of isolation to a different environment has a clear effect on the possibility of preserving a genetically fixed state.

\subsection{Inter-Agent Interactions}

Neighboring agents each have variable-states bundled within them.  In a complex agent, the number of variables is very large and in interacting agents, who are considered to be individuals, both similar and dissimilar variables leading to possible optimal and non-optimal alignments and collaborations must be present. In some cases, the non-optimal interactions can be turned off between two agents who are interacting based on another pair. Then one agent can then shuffle his neighbors and choose another agent who has a more favorable alignment along the previously {\it turned off} variable to interact with based on that variable. Hence we can consider a statistical ensemble of interactions among agents in a manner similar to interactions with variables spread in a local environment.

This may lead to an agent belonging multiple identity groups depending on different variables.  This scenario would imply involving some agents in one interest or collaboration group and others in other groups, and may be made possible by statistically distributing expressions of variables depending on their weights and coupling constants so that $<H_a>$ has a weighted optimal value when distributed over time, and only a subset of variables are active in segments of the time period.

The idea of a social cluster being partitioned into different coverings based on variables was discussed (Shafee and Steingo 2008).

As a possible scheme for inter-agent interactions given the bundled properties within each agent, and the option of not expressing all variables simultaneously, we can define subspaces within a cluster that are based on mutually important variables for subgroups.  These subgroups do not define a unique partition within the group, and one agent can belong to more than one such state-domains.  A simple model would involve
the cluster of agents being partitioned into {\bf m} weighted coverings for an agent showing {\bf m} variables.  Hence, he will have a subgroup of agents sharing the same variable-state interest, who can collaborate.  However, since each individual is different, partitions based on different variables might involve different subgroups for collaboration purposes.  Interesting dynamics may arise because of the spatial and temporal restriction for expression of collaborative projects and the number of different subgroups affiliated with the agent's idea of self based on similarity in a variable.  This situation is worth exploring in detail for different scenarios.

However, the notion of nested variables and the need for maintaining a critical minimum become important here, since if the total Hamiltonian for the agent exceeds the critical value at any time, the agent must perish. So optimization over any variables must have a feedback from the higher hierarchy level in the connected identity as the variable's own existence depends on preserving the identity of the agent at a higher hierarchy level, where couplings among variables are collaborative.

When both optimal and non-optimal interactions must be expressed within a shared environment of two agents, the weights of the variables leading to the value of the total interaction Hamiltonian becomes important in determining the relationship between the agents.

We can mathematically formulate the situation as follows:

Let us assume that two agents interact based on a certain variable (form friendship or animosity based on the value of $J_{ij}^a$) only when they can intercept the state of the other agent.

In the simple model, we also assume that two agents interacting within a shared environment are aware of each other's state along a certain variable, a, only when the states of both the agents are expressed.

Let the variable exist in two states, $S_i^{a_1}$ and $S_i^{a_2}$.  The corresponding common environment state can also come in $h^{a1}$ and $h^{a2}$.

Now, let us say agent i has $S_i^a$, which is aligned along the common environment, $h^a$.  Now we say that agent j has $S_i^{a'}$, which may or may not be along the state of agent i and the environment.

Given the limited number for expressible variables, we assume that each agent has a buffer state $S_i^{buff a/b}$, which can make the agent interact with the environment based on either variable a or variable b, so the agent may choose to express either a or b.  In the first case, the two agents are aware of each other's state in variable a, and in the second case, they are not.

Hence, when the variable states are expressed, we have the possible formulation for both agent's inter-agent and agent-environment terms:

\begin{eqnarray}
-J_{ij}^a S_i^a . S_i^{a'} - J_{ih}^a h^a . S_i^a \\
-J_{ij}^a S_i^a . S_i^{a'} - J_{jh}^{a'} h^a . S_i^{a'}
\end{eqnarray}

Here, the resultant interaction depends on whether or not the two states are optimal, which is known to both agents once they choose to express their states in the environment based on their own $J_{ih}^a$ and the alignment of their variable with the environment.

However, when the buffer state comes into play and both agents choose to express variable b instead of a, the relevant terms are

\begin{eqnarray}
0-  h^a [h^b . S_i^{buff_b} J_{ih}^b S_i^b S_i^a] \\
0-  h^a [h^b . S_j^{buff_b} J_{jh}^b S_j^b S_j^a]
\end{eqnarray}

In this case, although the agents do not interact to express their states in the environment because of the buffer state, the environment may interact with the agents, making them happy or unhappy depending on whether the environment state is optimal with the agent state or not.

In the other possible cases, one agent may choose to express its state in the environment, making his state clear to the second agent, who then chooses whether to express his state or not in a game for optimality.

\section{Accumulated Frustration}
In the history of social dynamics, revolutions can be seen as episodes of social reorganizations that take place as a result of accumulated frustration {}.  Again, in some individuals, often outbursts are seen after accumulated frustration.
We try to understand the two phenomena in terms of reorganization, non-optimized interactions and the bundled variable array.

In the first case, social reorganization requires large scale restructuring and re-configuration of sub-clusters, boundaries and tags, involving relearning and expenditure of energy in the process of relearning (Shafee 2004).  Such a process may involve the input of external energy from states in contact with the social system that is large enough to take the system to a more stable local minimum in a manner similar to spin-glass type quake.  This reorganization may be with respect to one non-optimized variable, such as race, gender, class or any other discerning variable that divides the cluster into subclusters.  The process might reorganize the sub-clusters into more stable configurations in that variable space (Shafee and Steingo 2008), often allowing for mixing of information among separated domains, and coming to a halt or decelerating at a point of local stability before complete reorganization takes place.

Again, the local minimum indicating decrease of frustration with respect to one variable may not make the cluster stable with respect to other variables.  However, the cluster might lack the existence of a record parameter with respect to a second variable, making the cluster itself unstable in a multidimensional space (Shafee and Steingo 2008).  In a heterogeneous cluster with sub-clusters possessing different skill levels, a stable minimum may be unreachable in a reasonable time period because of mismatch in variables in many dimensions so that optimizing in one dimension leaves another dimension non-optimized.  The existence of correlated variable sub-clusters together with conflicting preferences and histories may make it hard to create locally stable functioning systems with a workable labor hierarchy (Melvern 2004), especially if after the initial reorganization process, correlated variables start getting uncorrelated with different speeds (Shafee and Steingo 2008).  If such a population is confined to a shared physical space, so that agents are within close influence distances of conflicting agents, instability may erupt. Hence periods of unrest may prevail before a social cluster tries to reorganize itself if the cluster contains several conflicting sub-domains with scarce shared resources.  Again, the balance between individual variable satisfaction and security derived from forced collaboration in a heterogeneous society may be seen in spurts.  Examples such as the communist revolution (which itself resulted from cumulative frustration in procuring basic necessities, that was fulfilled at a mass level by forcing individuals to forgo personal freedom for the sake of a predefined good for the larger population) accumulating eventually to another large scale change resulting from dissatisfaction in expressing individual identities (Gorbachev 1988) or the outbreak of ethnic conflicts following sudden removal of forced clustering (Malcolm 1994).

In the case of personal accumulated frustration, the dimensionality of variable space and the existence of record parameters with respect to variables causing systems to jump to configurations where the interaction Hamiltonian is optimized with respect to one variable but not optimized in the overall connected variable-space may be used.  This may account for individual's needed personal space within a social cluster.  Hence, the Hamiltonian in a social spin glass type model may be created by coupling subsystems where each variable has an interaction Hamiltonian, but the variables themselves are coupled within the individual, leading to a total coupled Hamiltonian.  Delays processing the coupled Hamiltonian utilities or the existence of nested identity domains with different degrees of connectivity and coupling within an individual may account for competitions among variables and fluctuations in the direction of optimization of single variables and may be worth studying in detail.

On the other hand, an individual placed in an environment where one of his variable states is suppressed in a non-optimal internal configuration, the situation can be described by dissatisfaction.  With the progress of time, interactions and memories accumulated due to connectivity with a dissatisfied variable may enhance the fraction of total dissatisfaction.  Hence, we can add a term indicating total dissatisfaction because of the existence of an non-optimized variable.  When this dissatisfaction reaches a critical value, the variables undergo momentary reorganization to reset that level of dissatisfaction.

\begin{figure}[ht!]
\begin{center}
\includegraphics[width=10cm]{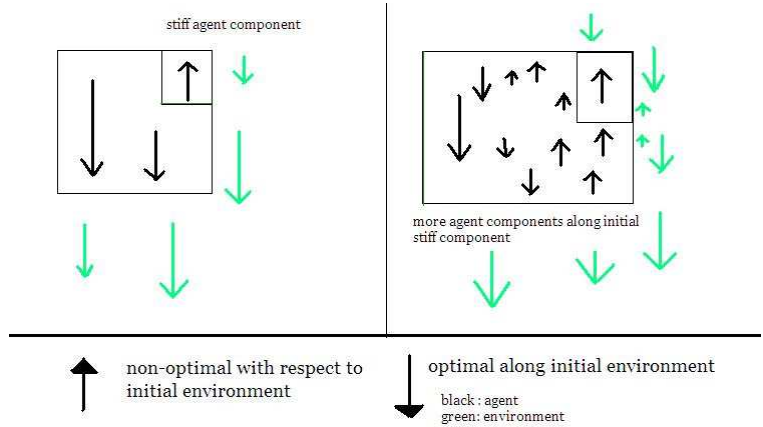}
\end{center}
\caption{\label{fig1} Optimal and non-optimal components within an agent competing to convert the other along it, causing frustration with respect to accumulated orientations and the orientation of the environment}
\end{figure}
This scenario, in the simplest case, can be model in the same manner continuous change with integrated J was formulated.  We use our concept of hierarchical interaction among components within the agent and the environment.  We assume that initially a portion of the system is non-optimal due to stiffness in shifting or competing interactions.  We start by dividing the components within the system into two broad interacting subsets, one optimal and another non-optimal, and we observe the evolution with time when different hierarchy components interact.
The relevant parts of the equation are:
\begin{eqnarray}
H(t)^m = [H_{opt}]^m \bigcup [H_{non-opt}]^m \\
H(t)' = -[h] . S_{opt}(t)^m[S_{opt}(t)^{m-p} . S_{non-opt}(t)^{m-p}]
S_{non_opt}(t)^m . [h]\\
\rightarrow H(t_1)= H(h(t_1), [S_{opt}]^m(t_1), [S_{opt}]^m(t_1),[S_{non-opt}]^m(t_1))
\end{eqnarray}

In the second expression, the nested interactions in the lower hierarchy levels are shown within the third parentheses.   It schematically shows that $S_{opt}(t)^m $ is made up of components of the type $S_{opt}(t)^{m-p}$, and $S_{non_opt}(t)^m$ is made up of $S_{non-opt}(t)^{m-p}$.  These components interact among themselves that later may change the higher level components. From now on, one variable followed directly by another variable(s) within a third parenthesis without any dot will be used to show nested hierarchical components and interactions.

If $S_{non-opt}$ is stiff, the time evolution will make the non-optimal part grow since interaction with neighboring lower hierarchy terms will keep the non-shiftable component steady, changing smaller neighboring components to align along it.

After the non-optimal component grows to exceed a threshold so that the system now has two large competing subsystems that must stay together if the system is to be viable, reorganization might be needed.  The new dynamics involves noting that it may be impossible to optimize with respect to all variables in a connected variable space, giving rise to jumps and fluctuations to semi-stable systems that may be less optimized in an overall variable space, and also accumulated dissatisfaction for suppressed variables.  This may indicate a new type of memory in social systems that is different from spin glass systems. Although spin glass systems progress towards more and more stable configurations, the connectivity of many variables within an agent, and the fixed states within an agent may give rise to a memory of integrated dissatisfaction or non-optimization in one variable dimension.

\section{The Nested Hamiltonian}
We end our study of the possibility of using spin glass like interactions in social complexity models by examining identities expressed at different levels of hierarchy, and the nested interconnected nature of such identities.

We have already written down terms regarding first order interactions among variables between two agents (optimally aligned or non-optimal, corresponding to similarity or conflict).  We now look at second order terms that bring more intricate dynamics.

The first case is regarding the constraint of resources and space, and the effect of usage.  Such effects were graphed in (Shafee 2009).  We model these effect in a simple manner for by making the shared environment component a compound component that includes other agents attached to it.  A term

$[h^a A_i^a.......A_m^a]$ treats the agents $[A_x^y]$ attached to the environment component $h^y$ as part of the environment, causing macroscopic change in the total environment state.  Hence, when the agent interacts with the compound environment state, in order to align the environment, he must also counter the effects of the other agents that are parts of the environment now, and so that he tries to align the total compound state along his preference, which might involve dealing with effects of other agents.

Hence, we include an interaction term:

$[h^a A_i^a.......A_m^a][A_i[S_i^a]]$

which indicates the interaction of the {\bf i}th agent with the compound environment state corresponding to his own variable state $S_i^a$.

Now, in the other end of the spectrum, a realistic model deals with effects of dissipation and repair by interaction with the environment and other variable states available that can replenish the lost components in a cost effective way.

It will be interesting to note the role of complexity in this context. Although the basic building blocks such as electrons and protons are identical particles that can come in various states, the states of higher level macro variables come from the number of such particles organized in different clusters at various levels as permitted by permissible combinations yielding stable energy values.  While atoms still have a discrete number of neutrons and protons in each type, and any two atoms of the same material have the exact same structure, as one moves up the hierarchy level, variables
have stable energy bands, and in more complicated organizations, no two objects are exactly the same although they can be categorized by similar exhibited properties.

When the effect of dissipation is studied, it must be noted that any part of an object that is repaired is highly unlikely to have the exact same replacement, although in a coarse grained world, they may appear to be almost the same.

Keeping this in mind, we model the effect of dissipation in our interaction model by

$A_i[S_i^A[V_a]h].O[V'^a]$

Here, the whole term left to the dot signifies the i'th agent who is a function of the variable state $S_i^a$ that itself has a sub-structure, $[V^a]$, which is also interacting with the environment in a competing manner, and may be lost.  The right side of the dot indicates an object, O, which h
as a somewhat similar variable $V'^a$ that can replace a lost variable $V^a$.

We have used two examples about how the hierarchical notion of identities in a complex system come into play in an interaction based model.  Such interactions can be extended to cases when identities are not the agents, but clusters, that contain agents.

\section{Future Extensions and Conclusion}
In this paper, we have discussed how certain aspects of a complex hierarchical network can be integrated into a spin-glass type to predict the behavior and dynamics of such systems.

We have indicated how several concepts specific to complex systems that exhibit meta-stable nested identity type sub-groups can be incorporated into an interaction-based system with a narrow range of constraints to allow for the emergence of such behavior.  We have also discussed how stability can be seen from the viewpoint of minima in an energy landscape.

The notion of complexity was explored in regard with the number of possible expressible variables in a diverse complex system and the number of variables allowed to be expressed because of the constraints imposed by the system's physical realizations and topology was restricted.  It was also stated how statistical expression of variables within an agent's environment can be indicative of weights of such variables.
However, as an agent interacts with the environment and modifies itself, such weights can change.  This can be very similar to change of weights in neural networks, where repeated interactions can make neurons change configuration slightly.

This paper discusses the formulation of behavior and the formation of groups based on interaction among favorable variable-states.  However, as was discussed, all variable-states of two agents are not expressed in the environment at the same time.
Hence, two agents choosing to interact must somehow predict the weighted variable-state array of the other agent.  This is a game of incomplete information.  Learning is based on past interactions and statistical expression of states of the other agent from past knowledge.  Formulating a prediction algorithm is an interesting problem.

In human cognitive processes, reactions are caused by cues evoked from past experiences (Raaijmakers and Shiffrin 1980), or by means of in-built reflex actions programmed by evolution.
Besides well-thought sequences of decisions, actions are also provoked by strong emotions related to an event can lead to increased production of chemicals.
In the mechanism involving cues, segments of history exist as correlated events within the brain so that when neurons corresponding to one event fire, it also stimulates another set of neurons corresponding to a second event.

When the second event is related to a strong emotion, the first one also evokes the second one, and hence a similar reaction.  Mechanisms of correlating environmental histories also plays a role in behavior of a complex system.  Repeated occurrence of a pair of events imprints the two to as highly correlated (Raaijmakers and Shiffrin 1980).
However, the programming of instant reactions based on correlation among variables or events in one hand brings in stability in the complex system by predicting the most probable quickly, but also makes the system averse from discarding two similar data points that are actually not connected, and are parts of longer strings that are not followed for time constraints.

The evolution of a complex system that has too many connected variables to be expressed or calculated at the same time, must make a balance between programmed correlation and individual examination.  The length of sequence in correlated points is also interesting to study in a complex system where similar patterns repeat, and some of these patterns may be highly correlated but given the arrow of time, no infinite series can be predicted.

In this paper, we have laid the preliminary theoretical foundations for constructing an interaction based model involving nested identity states within a permissible environment.  We have discussed the effects of having many possible variables and states than can be expressed or predicted, and have discussed the possibility of assigning weights.  Future work must understand the process of prediction and the balance between in-built reflexes, emotions leading to actions and rational choices in the stability and dynamics of such systems in a realistic world.

\section*{Acknowledgement} The author is thankful to Kevin Mitchell for technical help.

\section*{Bibliography}

Anderson PE, Jensen HJ, Oliviera LP, Sibani P, Evolution in Complex Systems. Complexity. 2004;  10:49–56.

Beloff N, Yugoslavia: An Avoidable War.  New European Publications; 1997.

Berlekamp ER, Conway J,Horton J, Guy RK, Winning Ways for your Mathematical Plays (2nd ed.) A K Peters Ltd; 2001.

Black F, Scholes M, The Pricing of Options and Corporate Liabilities. J Pol Econ. 1973;  81(3):  637-654.

Bowles S  and  Gintis H, Persistent Parochialism: Trust and Exclusion in Ethnic Networks.  J. Econ. Behav. and Organiz. 2004;  55: 1-23.

Bryngelson  JD and Wolynes PG, Spin glasses and the statistical mechanics of protein folding. PNAS. 1987; 84(21): 7524-7528

Cumings B, The Two Koreas. New York: Foreign Policy Association; 1984.

Dall  J and Sibani P, Exploring valleys of aging systems: the spin glass case. Eur. Phys. J. B. 2003;  36: 233-243.

Darwin C, The Origin of Species. London: John Murray; 1981.

Edwards S and Anderson P, Theory of spin glasses. J Phys F. 1975; 5:965-974.

Eldredge N, and Gould  SJ, Punctuated equilibria: an alternative to phyletic gradualism. In T.J.M. Schopf, ed., Models in Paleobiology. San Francisco: Freeman Cooper. 1972;  82-115.

Ereshefsky M, The Poverty of the Linnaean Hierarchy: A Philosophical Study of Biological Taxonomy. Cambridge University Press: Cambridge; 2000.

Gell-Mann M, The Quark and the Jaguar: Adventures in the Simple and the Complex, W. H. Freeman: New York; 1995.

Gorbachev M, Perestroika: New Thinking for Our Country and the World, Harper and Row; 1988.

Hopfield JJ, Herz AZM, Rapid local synchronization of action potentials: toward computation with coupled integrate-and-fireneurons. Proc Natl Acad Sci. 1995; 92: 6655.

Kawachia  K, Sekia  CM, Yoshidab H,  A rumor transmission model with various contact interactions. J Theor Bio 2008; 253(1): 55-60.

Malcolm N, Bosnia A Short History. New York University Press; 1994.

Mayr E, Change of genetic environment and evolution. In J. Huxley, A. C. Hardy and E. B. Ford. Evolution as a Process. London: Allen and Unwin; 1954.

Melvern L, Conspiracy to Murder: The Rwandan Genocide. New York: Verso; 2004.

Newman MEJ, Barabsi AL, Watts DJ, The Structure and Dynamics of Networks. Princeton (NJ): Princeton Univ Press; 2006.

Pospisil HBD, Valcrcel J, Reich J, Bork P,  Alternative splicing and genome complexity. Nature Genetics; 2001; 30: 2930.

Raaijmakers JGW and Shiffrin  RM, SAM: A Theory of Probabilistic Search in Associative Memory. In: Bower GH, editor. The Psych of Learning and Motivation: Advances in Research and Theor. New York: Academic Press;.1980: 207-262.

Reichardt J and White D R, Role Models for Complex Networks. European Physical Journal B. 2007; 60: 217-224.

Shafee F, A Spin Glass Model of Human Logic Systems. xxx.lanl.gov. arXiv:nlin/0211013; 2002.

Shafee F, Chaos and Annealing in Social Networks.  xxx.lanl.gov. arXiv:cond-mat/0401191; 2004.

Shafee F, Spin-glass-like Dynamics of Social Networks. arXiv:physics/0506161v1; 2005.

Shafee F, Oligo-parametric Hierarchical Structure of Complex System. NeuroQuantology J. 2007; 5(1):85-99.

Shafee F, Lambert function and a new non-extensive form of entropy. IMA J of App Math. 2007b; 72, 785-800.

Shafee F and Steingo G, Bifurcation of Identities and the Physics of Transition Economies. Transition Economies: 21st Century Issues and Challenges ed Gergõ M. Lakatos. Nova Publishers; 2008.

Shafee F, Interactions Among Agent Variables and Evolution of Social Clusters. submitted; 2009.

Shafee F, Agent Components and the Emergence of Altruism in Social Interaction Networks.  arXiv:0901.3772v1; 2009b.

Shafee F, A Network of Perceptions. Neuroquantology. 2009c; 7:2.

Shafee F, The Problem of Quantum Measurement and Entangled States: Interactions Among States and the Formation of Detector Images. submitted; 2009d.

Sibani P, and Jensen HJ, How a spin glass remembers; memory and rejuvenation from intermittency data: an analysis of temperature shifts.  J. Stat. Mech. 2004; P10013.


Sherrington and Kirkpatrick SK, Solvable Model of Spin Glass. Phys Rev Lett. 1975; 35: 1792-1796.

Vincent E, Dupuis V, Alba M, Hammann J, Bouchaud JP, Aging phenomena in spin-glass and ferromagnetic phases: Domain growth and wall dynamics. Europhys. Lett. 2000; 50 (5): 674

von Neumann J, The Theory of Self-reproducing Automata. A. Burks, ed., Univ. of Illinois Press, Urbana, IL.; 1966.

Wolfram S, A New Kind of Science. Wolfram Media, Inc.; 2002.

\end{document}